\documentclass[usenatbib]{mn2e}
\usepackage{times}
\usepackage{epsfig}
\usepackage{colordvi}

\usepackage{graphicx}
\usepackage{amssymb}

\def\Rs{R_{\rm S}}
\def\Rin{R_{\rm in}}

\def\Rsph{R_{\rm sp}}
\def\rsph{r_{\rm sp}}
\def\msun{{\rm M}_{\odot}}
\def\rphin{r_{\rm ph,in}}
\def\rphout{r_{\rm ph}}

\def\Mdot{\dot{M}}
\def\Mdoto{\dot{M}_0}
\def\mdot{\dot{m}}
\def\mdoto{\dot{m}_{0}}
\def\mdotin{\dot{m}_{\rm in}}

\def\Mdotw{\dot{M}_{\rm w}}
\def \epsilonw{\epsilon_{\rm w}}

\def\Mdotedd{\dot{M}_{\rm Edd}}
\def\Ledd{L_{\rm Edd}}
\def\Lbol{L_{\rm bol}}

\def\Qrad{Q_{\rm rad}}
\def\Qadv{Q_{\rm adv}}

\def\erg{{\rm erg}}
\def\secinv{{\rm s}^{-1}}

\def\Tph{T_{\rm ph}}
\def\Tsph{T_{\rm sp}}
\def\Tcsph{T_{\rm c,sp}}
\def\Tmax{T_{\rm max}}
\def\Tcmax{T_{\rm c,max}}
\def\Tphin{T_{\rm ph, in}}
\def\Tphout{T_{\rm ph}}

\def\Temp{T}

\def\vk{v_{\rm K}}
\def\omegak{\omega_{\rm K}}

\def\fc{f_{\rm c}}
\def\sigmasb{\sigma_{\rm SB}}

\def\rmd{{\rm d}}

\newcommand{\be}{\begin{equation}}
\newcommand{\ee}{\end{equation}}
\newcommand{\beq}{\begin{eqnarray}}
\newcommand{\eeq}{\end{eqnarray}}

\def\apj{ApJ}%
\def\apjl{ApJ}%
\def\aap{A\&A}%
\def\mnras{MNRAS}%
\def\pasj{PASJ}%

\title[On the nature of ultraluminous X-ray sources]
{Super-critically accreting stellar-mass black holes as
ultraluminous X-ray sources}

\author[J. Poutanen et al.]
   {Juri Poutanen,$^{1,2}$\thanks{E-mail: juri.poutanen@oulu.fi  (JP); galja@sai.msu.ru (GL); fabrika@sao.ru (SF)}
  Galina Lipunova,$^3$
  Sergei Fabrika,$^{4}$\footnotemark[1] 
  Alexey ~G.~Butkevich$^{1,5}$ \newauthor
  and Pavel Abolmasov$^{4}$\\
$^1$Astronomy Division, P.O. Box 3000,
           90014 University of Oulu, Finland\\
 $^2$KIPAC, Stanford University
P.O. Box 20450, MS29, Stanford, CA 94309, USA \\
$^3$Sternberg Astronomical Institute, Moscow State University, Universitetskij pr. 13, 
119992 Moscow, Russia \\
$^4$Special Astrophysical Observatory, 369167 Nizhnij Arkhyz, Karachaevo-Cherkesiya, Russia\\
$^5$Pulkovo Observatory, Pulkovskoye shosse 65,
          196140 Saint-Petersburg, Russia}

\begin{document}

\date{Accepted 2007 February 22. Received 2007 February 19; in original form 2006 July 28}
\pagerange{\pageref{firstpage}--\pageref{lastpage}} \pubyear{2007}
\maketitle
 
\label{firstpage}

\begin{abstract}
We derive the luminosity-temperature relation for the super-critically 
accreting black holes (BHs) and compare it to the data on ultraluminous X-ray sources (ULXs). 
At super-Eddington accretion rates, an outflow forms within the spherization radius. 
We construct the accretion disc model accounting for the advection and the outflow, 
and compute characteristic disc temperatures. 
The bolometric luminosity exceeds the Eddington luminosity $\Ledd$ by a logarithmic factor 
$1+0.6\ln\mdot$ (where $\mdot$ is the accretion rate in Eddington units) 
and the wind kinetic luminosity is close to $\Ledd$.
The apparent luminosity  for the  face-on observer is 2--7 times higher because of geometrical beaming.
Such an  observer has a direct view of the inner hot accretion disc, which
has a peak temperature  $\Tmax$ of a few keV in stellar-mass BHs. 
The emitted spectrum extends as a power-law $F_E\propto E^{-1}$ down to the 
temperature at the spherization radius  $\Tsph\approx \mdot^{-1/2}$ keV.
We associate $\Tmax$ with a few keV spectral components and $\Tsph$ with the soft, 
0.1--0.2 keV components observed in ULXs.
An edge-on observer sees only the soft emission from the extended envelope,
with the photosphere radius exceeding the spherization radius by orders of magnitude.
The dependence of the photosphere temperature on luminosity is consistent with that
observed in the super-Eddington accreting BHs  SS~433 and V4641~Sgr. 
Strong outflows combined with the large intrinsic X-ray luminosity of the central BH
explain naturally the presence of the photoionized nebulae around ULXs.
An excellent agreement between the model and the observational data strongly argues   
in favour  of ULXs being  super-critically accreting, 
stellar-mass BHs similar to SS~433, but viewed close to the symmetry axis.
\end{abstract}

\section{Introduction}

A large number of ultraluminous X-ray sources (ULXs)   has been discovered 
in the nearby star-forming galaxies \citep[see][ for a review]{M04}. 
Their luminosities exceed considerably $2\times 10^{39}\erg\ \secinv$, the Eddington limit 
for a stellar-mass  ($\sim$10$\msun$) black hole (BH).
This was argued to be the evidence for existence of the 
intermediate-mass BHs  (IMBHs, with $M=10^2$--$10^4\msun$; \citealt{CM99}).
The large apparent luminosities  can also be produced by super-critical accretion 
on to a stellar-mass BH  
(\citealt{SS73}, hereafter SS73; \citealt{JAP80,A88,Lip99,Fab04,BKP06}), 
or by the geometric \citep{FM01,K01} or relativistic 
\citep{RLF97, KFM02} collimation of radiation.

A strong argument in favour of IMBHs is the presence of a soft,
0.1--0.2 keV component in their spectra  \citep{KCPZ03,MF03,MF04a}. 
Arguments against  the IMBH interpretation include
theoretical problems with their formation \citep{K01} and their non-standard spectra, 
which show a cutoff at a few keV \citep{SRW06}, while other BHs, stellar as well as super-massive, 
at a few per cent of Eddington luminosity have hard power-law-like spectra \citep{Z97}. 
Furthermore, the observed anti-correlation between  luminosity and temperature contradicts the 
$L\propto T^4$ law expected for standard accretion discs \citep{FK07} and
some ULXs show  a harder, 1--4 keV thermal component with the corresponding radius of 
only 30--40 km \citep{M00,SRW06}. All this raises further doubts  on the IMBH interpretation.
The soft components can be interpreted as 
signatures of an extended photosphere (SS73; \citealt{Lip99,KP03}), but 
then we need to explain the simultaneous presence of the harder emission. 

The low-frequency, 0.02--0.2 Hz, quasi-periodic oscillations detected from 
ULXs in M82 \citep{SM03}, Holmberg IX \citep{DGR06}, and NGC 5408 \citep{SMW07},  
if interpreted  as the Keplerian frequencies at the innermost stable orbit around a BH,
argue in favour of IMBHs. 
However,   oscillations with very similar frequencies 
have been observed from the BHs in the Milky Way, 
Cygnus X-1 \citep{VCG94a} and GRS 1915+105 \citep{MRG97}, which certainly are not IMBHs.

Important clues on the nature of ULXs come from the presence of extended 
photoionized nebulae around them \citep{W02,PM02, KWZ04}.
The observations imply a rather  isotropic source of  the ionizing radiation rejecting   
the idea of a strong beaming of radiation. 
The nebulae are dynamically perturbed with the velocity gradients  
of $\sim$ 50--100 km $\secinv$ on the scale of 50--100 pc  \citep{LBF05,FA06,RWG06}. 
This points towards the activity of the central engine in form of a wind or a  jet.
The ULX nebulae are  similar to W50, the nebulae around  SS 433, 
the only known persistent super-critical accretor in the Milky Way,  
radiating presumably around $10^{40} \erg\ \secinv$ in the UV \citep{DF97}.
A super-critically accreting compact source can produce strong winds and jets which 
can inflate the nebulae \citep{B80,LBF05,PGM06,FA06,AFSA07}.   
On the other hand, there is no good  physical reason why a sub-critically 
accreting IMBH would produce such a strong outflow.

The observed similarities between W50 and ULX nebulae 
lead us to consider seriously the idea that the central engines of ULXs are 
super-critically accreting stellar mass BHs similar to SS 433.  
We do not see directly the X-ray source in SS 433, but if observed   
along the symmetry axis, it  would be a bright X-ray source  \citep{Katz86}, 
which we interpret as an ULX \citep{FM01,K02,Fab04,BKP06}.

In this paper, we develop a model for the super-critical accretion disc 
accounting for the effects of advection and outflows.
We also construct a one-dimensional, vertically integrated model of the wind,
estimate the optical depth through the wind and determine 
characteristic temperatures as a function of the mass accretion rate and the luminosity.
We further compare the resulting  luminosity-temperature relations to the data on 
ULXs as well as the BHs in our galaxy and LMC.

\section{Subcritical accretion discs}
\label{sec:ss73}

The standard accretion disc  theory (SS73) can be applied when the 
accretion rate is not very high and 
the luminosity does not exceed  the Eddington limit 
\be 
\Ledd  =  \frac{GM\Mdotedd}{2\Rin} = \frac{4\pi GMc}{\kappa} = 
1.5 \ 10^{38} \ m \frac{1.7}{1+X}   \ \erg\ \secinv,
\ee
where $\Mdotedd=48\pi GM/c\kappa=2\ 10^{18}m \ \mbox{g} \ \secinv$ 
is the Eddington accretion rate, $\Rin=3\Rs$ is the inner disc radius,  
$\Rs=2GM/c^2$ is the Schwarzschild radius. 
The stellar mass, measured in solar masses, is $m=M/\msun$, 
$\kappa=0.2 (1+X)=0.34\  \mbox{cm}^2 \ \mbox{g}^{-1}$ 
is the Thomson opacity and $X$ is the hydrogen mass fraction 
(which we assume equal to solar $X=0.7$). 

The energy flux from one face of the disc (in Newtonian approximation)
as a function of radius $R$ is
\be  \label{eq:stand}
Q^+(R)=  \frac{3}{8\pi} \frac{GM \Mdot }{R^3} \left[ 1 - {r}^{-1/2} \right]  ,
\ee
where  $\Mdot$ is the accretion rate and $r=R/\Rin$.
This results in the effective temperature variation 
$T(R)\propto R^{-3/4}$ at large $R$ and the maximum  
observed color temperature 
\be \label{eq:tmax_ss}
\Tcmax   = 1.26 \fc \ m^{-1/4} \mdot^{1/4}  \ \mbox{keV} , 
\ee
which is reached at $r_{\max}=(3/2)^{4/5}\approx1.38$.
Here $\mdot=\Mdot/\Mdotedd$ is   dimensionless  accretion rate.
The color correction  $\fc \approx 1.7$   \citep[e.g.][]{ST95} 
describes the hardening of  the spectrum relative to the black body.
The emitted flux $F_E\propto E^{1/3}$ at $E\lesssim kT_{\rm c,\max}$. 
The total emitted luminosity, 
\be 
L=  \int_{\Rin}^{\infty}  Q^+(R) 4\pi R \rmd R =  \mdot \Ledd,
\ee
depends on the maximum temperature as  $L\propto \Temp_{\rm c,\max}^4$.

\section{Supercritical accretion discs} 

The matter supply to the BH  in a close binary 
can significantly exceed the Eddington rate, $\mdoto=\Mdoto/ \Mdotedd\gg1$. 
There are two different views on the way the accretion  proceeds in this regime. 

(i) The `Polish doughnut' or slim disc models \citep{JAP80,A88}
assume that all the supplied gas reaches the BH, but 
most of the gravitational energy released in the disc is advected into the hole as the 
photons are trapped in the flow.  
Inside the trapping radius $R_\mathrm{tr}$ (where photon diffusion and accretion
time-scale are equal),  the vertical component of gravity scales as $R^{-2}$ 
(as the disc relative scale-height $H/R$ is of the order unity)  resulting 
in the same radial dependence of the radiation flux $\Qrad$, because the disc 
is radiation pressure supported. 
This implies the effective temperature distribution $R^{-1/2}$ and 
a flat (in $EF_E$) emitted spectrum $F_E\propto E^{-1}$ \citep[see e.g.][]{WF00}.
Naturally, a logarithmic dependence of bolometric luminosity
of inner supercritical disc on $\Mdoto$ is yielded:
\be
L \propto \int\limits_{\Rin}^{R_\mathrm{tr}} \Qrad(R) R \ \rmd R
\, \propto  \,
\int\limits_{\Rin}^{R_\mathrm{tr}} R^{-1} \rmd R
= 
\ln\left(\frac{R_\mathrm{tr}}{\Rin}\right) \approx \ln \mdoto .
\ee
The last relation follows from the fact that the trapping radius 
scales with the accretion rate as
$R_\mathrm{tr} \approx \mdoto \Rin \gg \Rin$.
The total disc luminosity  exceeds the Eddington one
 by a logarithmic factor: 
\be
\Lbol  \approx \Ledd \ (1+\ln \mdoto) .
\ee
 
(ii) Alternatively, instead of spending most of the dissipated energy to increase 
the entropy of the gas, this energy can be spent to eject  the excess mass  (SS73). 
In this model most of the gas is blown away by the radiation pressure,
 the accretion rate decreased linearly with radius  $\dot M(R) \propto R$, 
and only a small fraction of it, $\mdot \sim 1$, makes it to the hole.\footnote{A similar physics 
might operate at a very low accretion rate, when the gas cannot 
cool resulting in a positive Bernoulli parameter. This implies that the gas 
is effectively unbound and can easily produce outflows \citep{NY94,BB99}.}
The linear dependence of the accretion rate on radius results in the  $R^{-2}$ 
dependence of the radiative flux.\footnote{We note here that any supercritical disc model (with advection, with outflows, or both) predicts such a dependence, which just follows from the vertical balance of gravity and radiative pressure.}
The luminosity and the disc effective temperature radial distribution are very similar 
to the slim disc case (see below). The largest difference is in the presence of 
a strong wind which blocks and reprocesses the radiation from the central part of the disc. 
The observed spectrum, therefore, depends on the velocity structure and geometry  
of the wind  and the position of the observer relative to the disc rotational axis. 

Reality might be somewhere in between, with both advection and outflow 
operating together as shown by the numerical simulations  \citep{ECK88,OM05,OTT05}.
Below we first construct a one-dimensional, vertically integrated 
model of the supercritical disc with the outflow and then
discuss the effect of the wind on the emergent spectrum. 

\subsection{Supercritical disc with outflow and without advection}
\label{sec:ssdisc_winds}

The scale-height of accretion discs  for $\mdoto \sim1$ 
is determined by the balance between the 
radiation pressure force and the vertical component of gravity (SS73), 
\be  \label{eq:scaleheight}
 \frac{H}{R} = \mdoto \frac{3}{r} \left[ 1 -  {r}^{-1/2} \right]  .
\ee
When the accretion rate exceeds the critical value $\mdot_{\rm cr} = 9/4$,  
the ratio $H/R$ exceeds unity at some radii    and the outflow is inevitable. 
Accurate calculations  \citep{BKB77} show that 
the outflow can start even at a smaller rate.
At $\mdoto \gg \mdot_{\rm cr}$, the disc 
starts `feeling' that it is supercritical at the spherization radius $\rsph=\Rsph/\Rin \approx \mdoto$,
where the outflow starts.

The wind affects the disc structure by removing the angular momentum. 
The modified angular momentum conservation equation is \citep[see e.g.][]{Lip99}
\be \label{eq:angmom} 
\frac{\rmd}{\rmd R}\left(   \Mdot(R) \omegak R^2 \right) = 
\frac{\rmd g(R)}{\rmd R} +  \omegak R^2 \frac{\rmd \Mdot(R)}{\rmd R} ,
\ee
where  $\omegak(R)=(GM/R^3)^{1/2}$ is the Keplerian angular velocity, 
\be 
g(R) = 2 \pi\ T_{r\phi} R^2 
\ee
 is the torque and $T_{r\phi}=2H\ t_{r\phi}$ is the vertically integrated viscous stress.
The first term  on the right hand side of equation (\ref{eq:angmom})
is responsible for transfer of the angular momentum by viscosity, while
the second term is the momentum transfer by the outflowing matter.
The viscous heating rate per unit area is 
\be\label{eq:qplus}
2 Q^+(R)=  - T_{r\phi} R \frac{\rmd \omegak}{\rmd R}= 
\frac{3}{2}T_{r\phi} \omegak  =  \frac{3}{4\pi} \omegak \frac{g(R)}{R^2}  .
\ee

If advection is not accounted for, all dissipated energy is converted locally 
to the radiation:
\be \label{eq:qplusqrad}
Q^+(R) = \Qrad(R) .
\ee
We can assume that a fraction $\epsilonw$ of the radiative energy goes to accelerate the outflow:
\be \label{eq:energy_adv} 
\epsilonw  Q_{\rm rad}(R) =  
\frac{1}{2} \frac{1}{2\pi R} \frac{\rmd \Mdot(R)}{\rmd R} \frac{\omegak^2 R^2}{2}  
= \frac{GM}{8\pi R^2}  \frac{\rmd \Mdot(R)}{\rmd R} .
\ee
where $\vk^2(R)/ 2=\omegak^2 R^2/2$ is the energy required to eject a unit mass 
from radius $R$ to infinity. 
\citet{Lip99} assumed the maximum outflow rate with $\epsilonw =1$ and 
obtained an analytical solution of the system of equations 
(\ref{eq:angmom})--(\ref{eq:energy_adv}), 
applying boundary conditions $\Mdot(\Rsph)= \Mdoto$ 
(where $\Rsph$ is a parameter of the model) and $g(\Rin)=0$:
\beq \label{eq:mdot}
\frac{\Mdot(r)}{\Mdotedd} & = &  \left\{  \begin{array}{ll} 
\displaystyle
\mdoto \frac{r}{\rsph} \frac{1+\frac{2}{3} r^{-5/2} }{1+\frac{2}{3} \rsph^{-5/2} } , & r \le \rsph , \\
 \mbox{ } &  \\
\mdoto , &    r > \rsph, 
 \end{array} \right. 
 \eeq
 \beq
\label{eq:torque}
\frac{g(r)}{g_0}         & = & \left\{  \begin{array}{ll} \displaystyle
\frac{\mdoto r^{3/2}}{3\rsph} \frac{1- r^{-5/2} }{1+\frac{2}{3} \rsph^{-5/2} } , 
&  r \le \rsph , \\  
\mbox{ } &  \\
\displaystyle
\frac{g(\rsph)}{g_0}   +\mdoto \left( {r }^{1/2}-{\rsph}^{1/2}\right) , &  r> \rsph .
 \end{array} \right.  
\eeq
where $g_0=\Mdotedd \sqrt{GM\Rin}$. 

At intermediate radii $1\ll r < \rsph$,  the accretion rate is close to the dependence 
$\Mdot(r) = \Mdoto\ r/\rsph$ suggested by SS73 and the radiative energy flux is 
\be \label{eq:qrad_w}
\Qrad(R) = \frac{3}{8\pi} \frac{(GM)^{1/2} g(r)}{R^{7/2}} 
 \approx  \frac{GM \Mdot(R)}{8\pi \ R^3}   \propto R^{-2} . 
\ee
Note that in  the standard model there is a factor of 3 present
in a similar formula (see equation \ref{eq:stand}), which is related to the fact
that the energy dissipated close to the inner disc radius is transported to larger 
radii and radiated there. In the model with the mass loss, this factor  is missing, 
implying that the energy transfer from the inner part of the disc is negligible. 
 
 The luminosities produced within and outside $\rsph$ are
 \beq \label{eq:lum}
 \frac{L(r< \rsph) }{\Ledd} & = & \frac{\mdoto}{\rsph} 
 \frac{\ln \rsph  - \frac{2}{5} \left( 1 -   \rsph^{-5/2}  \right) } {1+ \frac{2}{3}   \rsph^{-5/2} }  , \nonumber \\
\frac{L(r>\rsph) }{\Ledd}  & = & \frac{5}{3}  \  \frac{\mdoto}{\rsph} 
 \frac{1}{1+ \frac{2}{3}  \rsph^{-5/2} } .
 \eeq
Defining the spherization radius by the condition (SS73) 
\be \label{eq:lrsph}
L(r> \rsph) = \Ledd, 
\ee
 we obtain 
\be \label{eq:rsph}
\rsph \approx \frac{5}{3}\ \mdoto , 
\ee
which gives the total luminosity 
\be \label{eq:ltot}
L\approx \Ledd\left(1  +  \frac{3}{5} \ln  \mdoto \right)
\ee
and the Eddington accretion rate to the BH, $\mdot(\Rin)=1$.
A fraction of this luminosity can escape as radiation and the rest as kinetic energy of the 
outflow.

It is worth noting here 
that if we assume in this model that only a small fraction of the dissipated energy is used to 
produce the outflow, i.e. $\epsilonw<1$ 
(see eq. \ref{eq:energy_adv}), then the resulting accretion rate is too high.
With the high radiative efficiency (because we neglect advection)
the locally emitted flux is strongly super-Eddington, and therefore the model is unphysical.
However,  accounting for the advection allows us to construct a self-consistent model
even in this case (see below). 

\begin{figure*}
\begin{center}
\leavevmode \epsfxsize=7.5cm \epsfbox{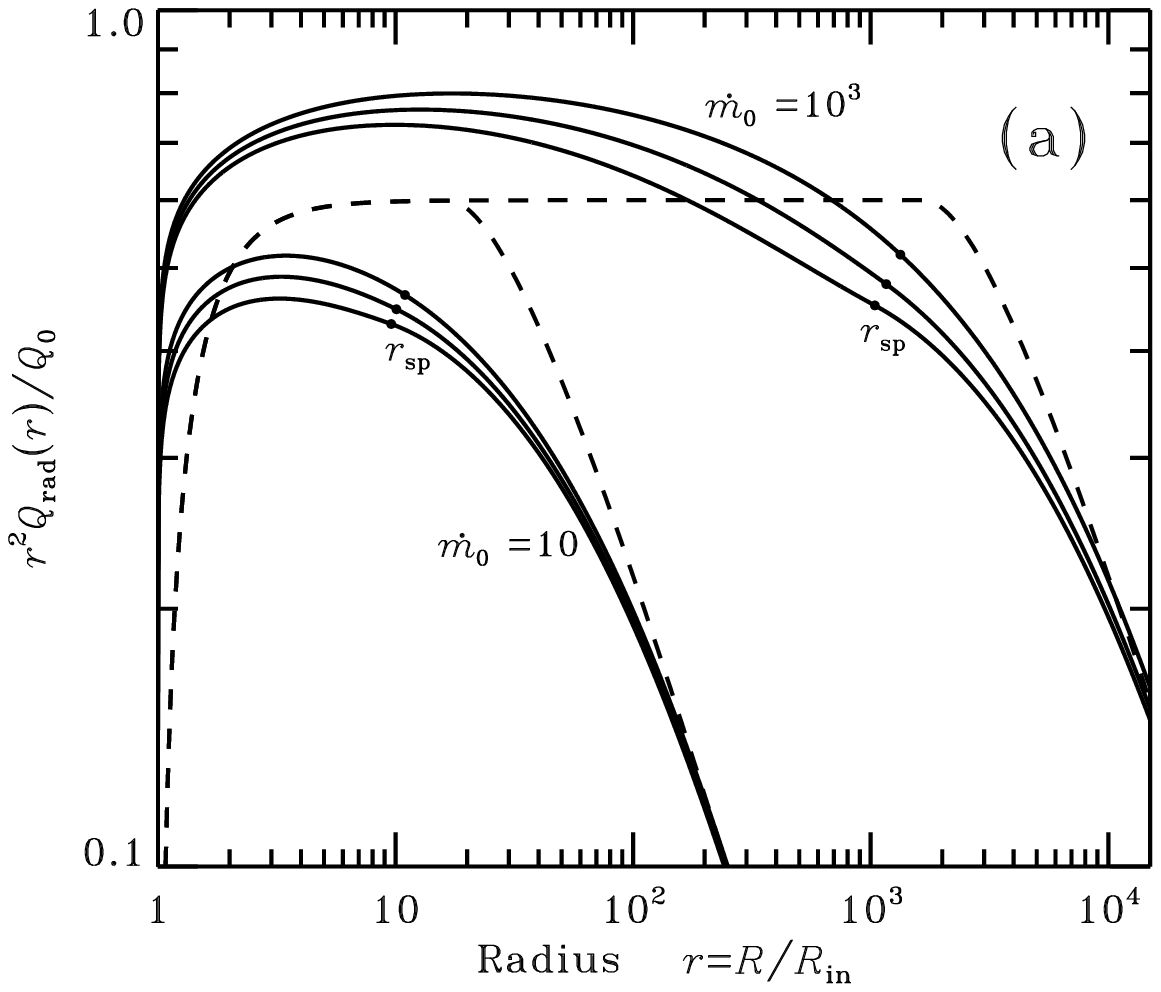} \hspace{1cm}
\epsfxsize=7.5cm \epsfbox{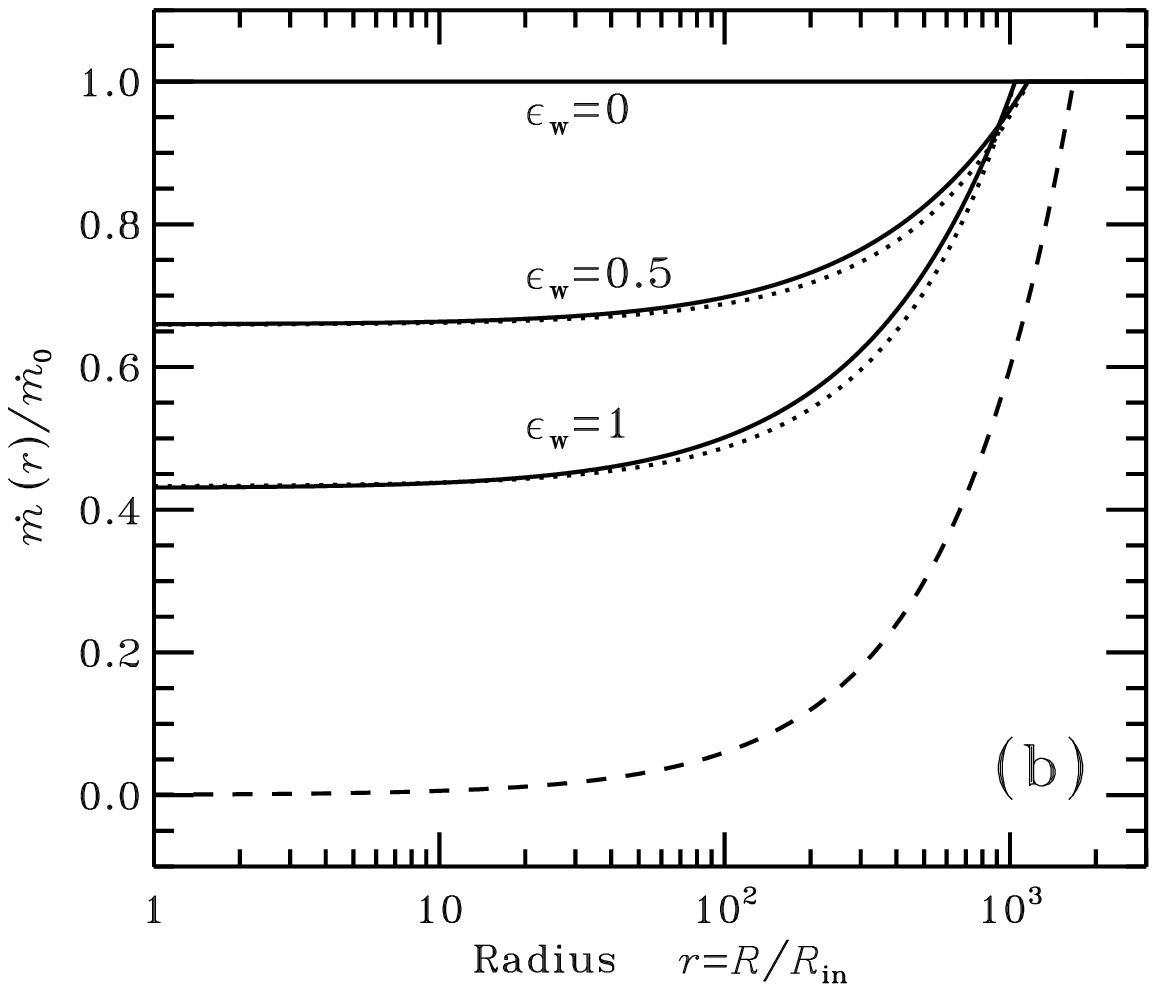}
\end{center}
\caption{ \label{fig:qradmdot}  
(a) The radiative energy flux emitted from the disc (times $r^2$, i.e. 
energy flux per logarithm of radius) 
as a function of radius for $\mdoto=10$ and $10^3$.
The flux is measured in units 
$Q_0= GM\Mdotedd/8\pi\Rin^3= c^5/36 GM \kappa =1.52\ 10^{25} m^{-1}  
\mbox{erg\ cm}^{-2} \ \mbox{s}^{-1}$. 
The solid curves from the top  to the bottom  are the exact solutions of the advective disc 
equations for $\epsilonw=0, 1/2$ and 1. 
Bold dots represent the positions of the spherization radius defined by 
condition (\ref{eq:lrsph}).
The dashed curves  correspond to the analytical solution given by equations 
(\ref{eq:qplus}), (\ref{eq:qplusqrad}) and (\ref{eq:torque}). 
 (b) The accretion rate as a function of radius for $\mdoto=10^3$.
The solid curves are the exact solutions for the advective disc, while 
the dashed curve is solution (\ref{eq:mdot}) for  the non-advective disc with the outflow.
The dotted curves are the linear approximations given by formula (\ref{eq:linapp_mdot}).
}
\end{figure*}

\subsection{Supercritical disc with advection}
\label{sec:advdisc_winds}

At supercritical accretion rates, the advective transport of the 
viscously generated heat to the black hole becomes important in the energy
balance equation. 
The diffusion time for photons, traveling to the disc surface, 
becomes larger than the characteristic time of the radial displacement of
the matter.  Due to  the advective removing of the heat,  the locally
radiated flux becomes smaller:
\be
\Qrad = Q^+ -  \Qadv \, .
\ee
A model of a supercritical advective disc
with an outflow was first proposed by \citet{Lip99} (see also
\citealt*{Kit02,Fuk04}). The maximum mass loss from an
advective disc by the energy-driven wind is about twice smaller 
than in the model without advection and amounts to about 3/5  of initial accretion
rate $\Mdoto$.
We modify the model considered by  \citet{Lip99} 
assuming that a fraction $\epsilonw<1$ of the radiation energy flux is 
spent on the production of the outflow (see eq. [\ref{eq:energy_adv}]). 
We solve the standard set of equations for the advective disc 
(as described in details in \citealt{Lip99}) for various  $\mdoto$ and $\epsilonw$.

The outflow occurs within the spherization radius $\rsph$, which 
is defined self-consistently from condition (\ref{eq:lrsph}).
For different $\epsilonw$, the spherization radius
varies because a different amount of angular
momentum is gone with the wind, and, consequently,  the structure of the outer
subcritical disc is different as it depends on the boundary condition at $\rsph$. 
The results of our calculations can be approximated (with the accuracy 
of 2 per cent  for $\mdoto>5$) by a simple formula:
 \be \label{eq:rsphapp}
\frac{ \rsph}{\mdoto} \approx  1.34 -0.4 \epsilonw+0.1 \epsilonw^2 - 
(1.1-0.7 \epsilonw ) \mdoto^{-2/3}. 
\ee

The resulting radiation flux and the mass accretion rate as functions of radius 
are shown in Fig.~\ref{fig:qradmdot}. 
 One sees that the radiative energy flux depends very weakly on $\epsilonw$. 
This also means that a similar solution will be obtained if a smaller mass 
 is ejected with a larger velocity. 
 The radial dependence of  $\Qrad (r)$ is similar to that in discs with strong mass 
loss and no advection (see section \ref{sec:ssdisc_winds} and the dashed curve 
in Fig. ~\ref{fig:qradmdot}a).
 At $r< \rsph\approx \mdoto$, $\Qrad (r) r^{2}$ is almost constant. 
 This is just the consequence of the fact that the disc is close to the Eddington limit locally 
 everywhere, i.e. $H/R$ is close to unity. 
At $r \gg \rsph$,  $\Qrad$ decreases as $r^{-3}$ as in the standard disc. 
The total luminosity is well approximated by equation (\ref{eq:ltot}).
As we assume that a fraction  $\epsilonw<1$ of the radiative energy flux  
is spent to drive the wind to infinity, a  fraction $1-\epsilonw$ can escape as radiation. 
We note that in the advective disc (with or without the outflow) the
scale-height (see \citealt{B98,Lip99}, and Fig. \ref{fig:qradmdot}a)
\be
H/R \approx \frac{r^2\ \Qrad(r)}{Q_0} < 0.8 ,  
\ee 
which justifies our use of the vertically integrated quantities.

Because the outflow is optically thick at most radii and the radiation 
is partially trapped (see below),  it is  {\it energy-driven} (not  {\it momentum-driven} 
as was assumed by \citealt{KP03}), and its kinetic luminosity  can exceed $\Ledd$.
 
The fraction of the initial accretion rate that passes through the inner radius can be approximated 
for $\mdoto>2.5$ as
\be  \label{eq:mdotinapp}
\frac{ \mdotin }{\mdoto}  \approx 
\frac{1-a}{1- a \left( \frac{2}{5}\mdoto  \right) ^{-1/2}  } , 
 \ee
where $a=  \epsilonw (0.83 - 0.25 \epsilonw)$.
 For example, for  $\mdoto=1000$ and $\epsilonw=1/2$ we get 
  $\rsph=1.16\ \mdoto$ and $\mdotin=0.66\ \mdoto$, while 
 in the case of the maximal outflow  with $\epsilonw=1$ we have 
  $\rsph=1.04\ \mdoto$ and $\mdotin=0.43\ \mdoto$.
 The total outflow rate for large $\mdoto$ is thus 
 \be 
 \Mdotw \approx a \Mdoto . 
 \ee
 
The exact solutions for the accretion rate inside the spherization radius 
 can be approximated by the linear relation 
 (compare solid and dotted curves  in Fig. \ref{fig:qradmdot}b)
  \be \label{eq:linapp_mdot}
\mdot(r)  \approx \mdotin 
+ \left( \mdoto -  \mdotin \right) \frac{r}{\rsph} .
 \ee
This can be easily understood from equation (\ref{eq:energy_adv}): 
the radial derivative of the accretion rate is proportional to $r^2 Q(r)$, 
which is almost constant at $r< \rsph$. 
The linear behaviour of the accretion rate on radius is also 
obtained in the numerical simulations \citep{OM05}.

 \section{Structure of the outflow}

\subsection{Optical depth through  the outflow}

The gas ejected from the accretion disc at cylindrical radius $R$ gains the  velocity   
perpendicular to the disc (in the $z$ direction), 
$v_{z} = \xi \vk(R)$ with $\xi\gtrsim 1$. 
A mass element reaches asymptotically the  velocity of $\zeta \vk(R)$, 
with $\zeta=\sqrt{\xi^2-1}$. 
If the angular momentum is conserved, the gas
moves (in the ballistic approximation) along the line $z/R=\zeta$ at large radii. 
The radial (i.e. projected to the disc plane) velocity  is thus $\zeta \vk(R)/\xi$. 
Because the disc scale-height $H/R$$\sim$0.6 and 
$\xi$ is not expected to exceed 1.5--2 (owing to the energy constraints),
$\zeta$$\sim$0.6--1.7. 
The outflow is thus confined in the region outside the cone of 
opening angle $\theta$ given by $\cot \theta = \zeta$ and it occupies 
$\Omega_{\rm w} /4\pi= \cos\theta=$0.5--0.85 fraction of the sky.

 Let us construct the vertically averaged wind model.   
As our baseline we take the advective disc model with the outflow described in the 
previous section. 
We take the accretion rate as given by equation (\ref{eq:linapp_mdot}) with 
$\rsph$ and $\mdotin$ given by equations (\ref{eq:rsphapp}) and (\ref{eq:mdotinapp}). 
The mass outflow rate (at $r<\rsph$) is then 
\be 
\Mdotw(R)= \int_{\Rin}^{R} \frac{\rmd \Mdot(R)}{\rmd R}\rmd R =  
\Mdotedd (\mdoto-\mdotin) \frac{r-1}{\rsph} . 
\ee
Within the spherization radius 
the mean wind radial velocity should scale with the local Keplerian velocity. 
At $R>\Rsph$, the mass loss rate is constant and the gas reaches
asymptotically  velocity of $\sim \beta\ \vk(\Rsph)$,  where $\beta\sim \zeta \sim1$.
We can then approximate the wind radial velocity profile by a simple  function: 
\be \label{eq:w_velo}
 v_{\rm w} (R) =  \left\{ \begin{array}{ll}
\beta \sqrt{\frac{GM}{R} } , & R \le \Rsph, \\
 \beta\sqrt{\frac{GM}{\Rsph} } ,  & R> \Rsph .
  \end{array} \right.  
\ee

\begin{figure}
\centerline{ \epsfig{file=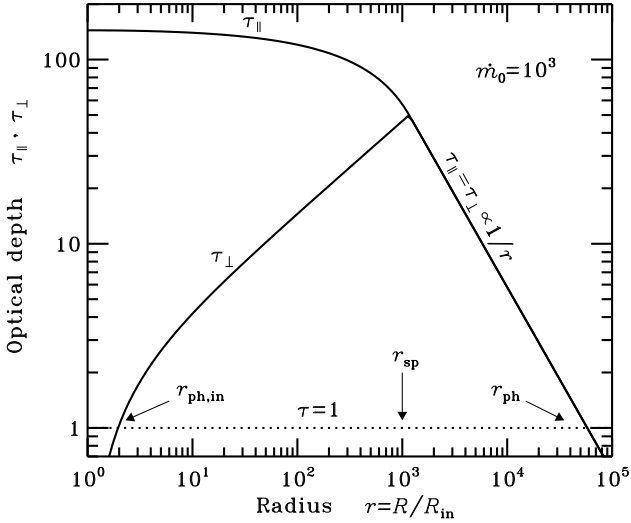,width=8.3cm}}
\caption{\label{fig:tau}  
The Thomson optical depth through the wind as a function of radius in the direction 
parallel and  perpendicular to the disc for $\mdoto=10^3$.  
The wind parameters are $\beta=\zeta=1$  and $\epsilonw=1/2$.
The perpendicular  optical depth reaches the maximum of 
$\sim$$3 \mdoto^{1/2} \frac{\epsilonw}{\beta}$$\sim$50 
at the spherization radius $\rsph \approx \mdoto$.
}
\end{figure}
 
From the mass conservation law,
\be 
\Mdotw(R)=  \zeta \  4\pi R^2 \ \rho(R) \ v_{\rm w} ,
\ee
we find the  mean density $\rho$ and the optical depth in the perpendicular direction, 
$\tau_{\perp} (R)=  \kappa {\rho} \zeta R$:
\be \label{eq:tauperp} 
\tau_{\perp} (r) = \frac{\tau_0}{\beta}  \frac{\mdoto-\mdotin}{\rsph} 
\left\{ \begin{array}{ll}
\displaystyle
   r^{1/2}  - r^{-1/2}    , &  r \le \rsph , \\ 
   & \\
    (\rsph -1) \rsph ^{1/2} r^{-1}  ,  & r > \rsph .\\
 \end{array} \right.  
 \ee
 Here $\tau_0$=$\Mdotedd \sqrt{6} \kappa/4\pi c\Rin\approx$5.
The optical depth  has the maximum 
$\tau_{\perp,\max}$$\approx$$ 3(\mdoto^{1/2}$$-$$\mdoto^{-1/2} ) \epsilonw/\beta$ at  $r=\rsph$ 
(see~Fig.~\ref{fig:tau}).

The Thomson optical depth from radius $r$ 
in the direction parallel to the disc is
\be 
\tau_{\parallel}(r)=  \kappa \int_R^{\infty} {\rho}(R')\rmd R' = \frac{1}{\zeta}
 \int_r^{\infty}  \tau_{\perp} (r') \frac{\rmd r'}{r'} . 
\ee 
The maximum  of $\sim 8(\mdoto^{1/2}-4/3) \epsilonw /\zeta\beta$ 
is reached at $r=1$. At $r>\rsph$ it decays in the way identical to $\tau_{\perp}$ (Fig.~\ref{fig:tau}):
\be \label{eq:taupar1} 
\tau_{\parallel}(r)  = 
 \tau_{\perp} (r)/ \zeta \propto r^{-1}.
\ee
The outflow becomes optically thick at $\mdoto\sim2.5$ for $\epsilonw \sim1/2$.

\begin{figure*}
\begin{center}
\leavevmode \epsfxsize=8.5cm \epsfbox{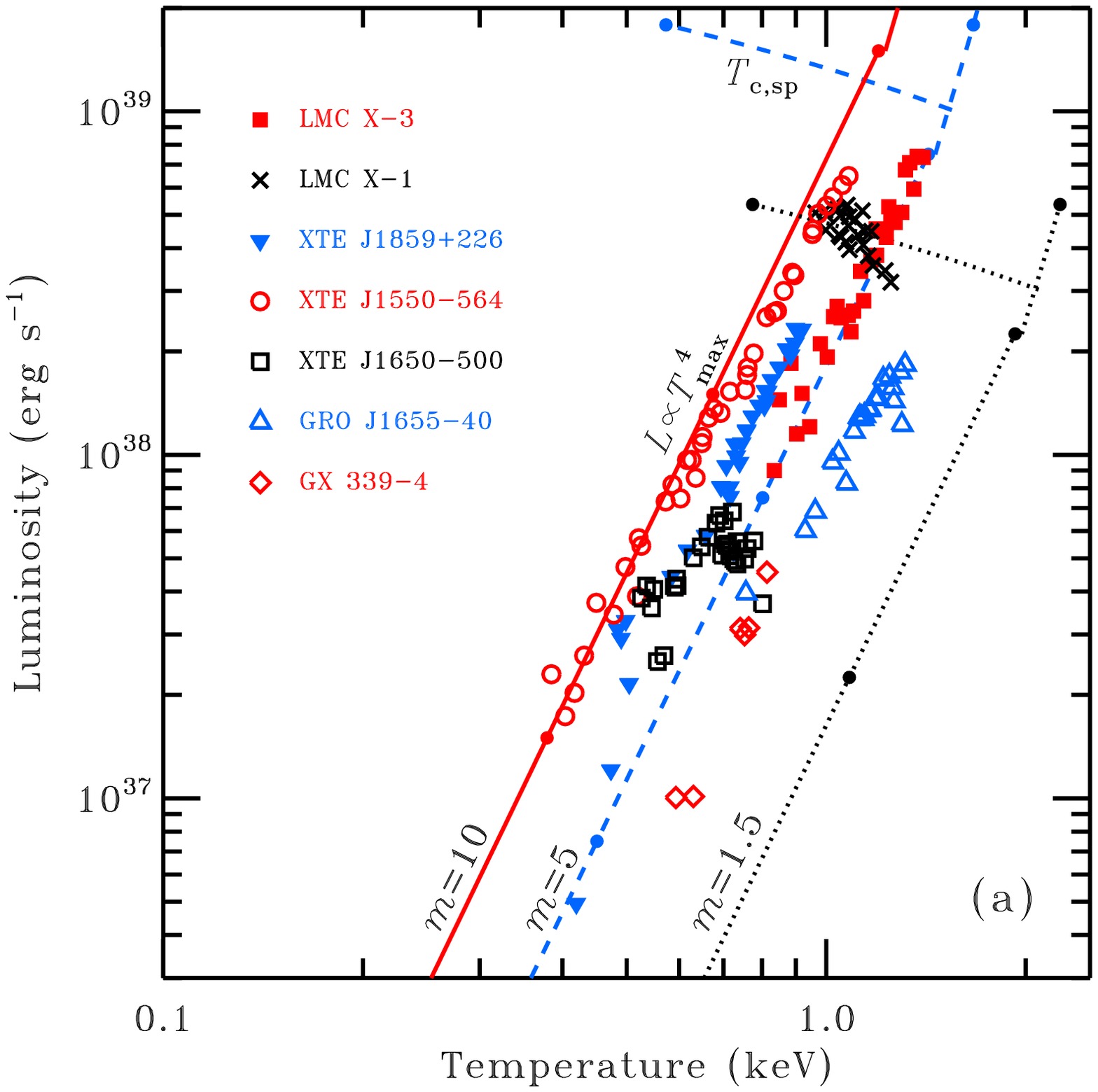} \hspace{0.4cm}
\epsfxsize=8.5cm \epsfbox{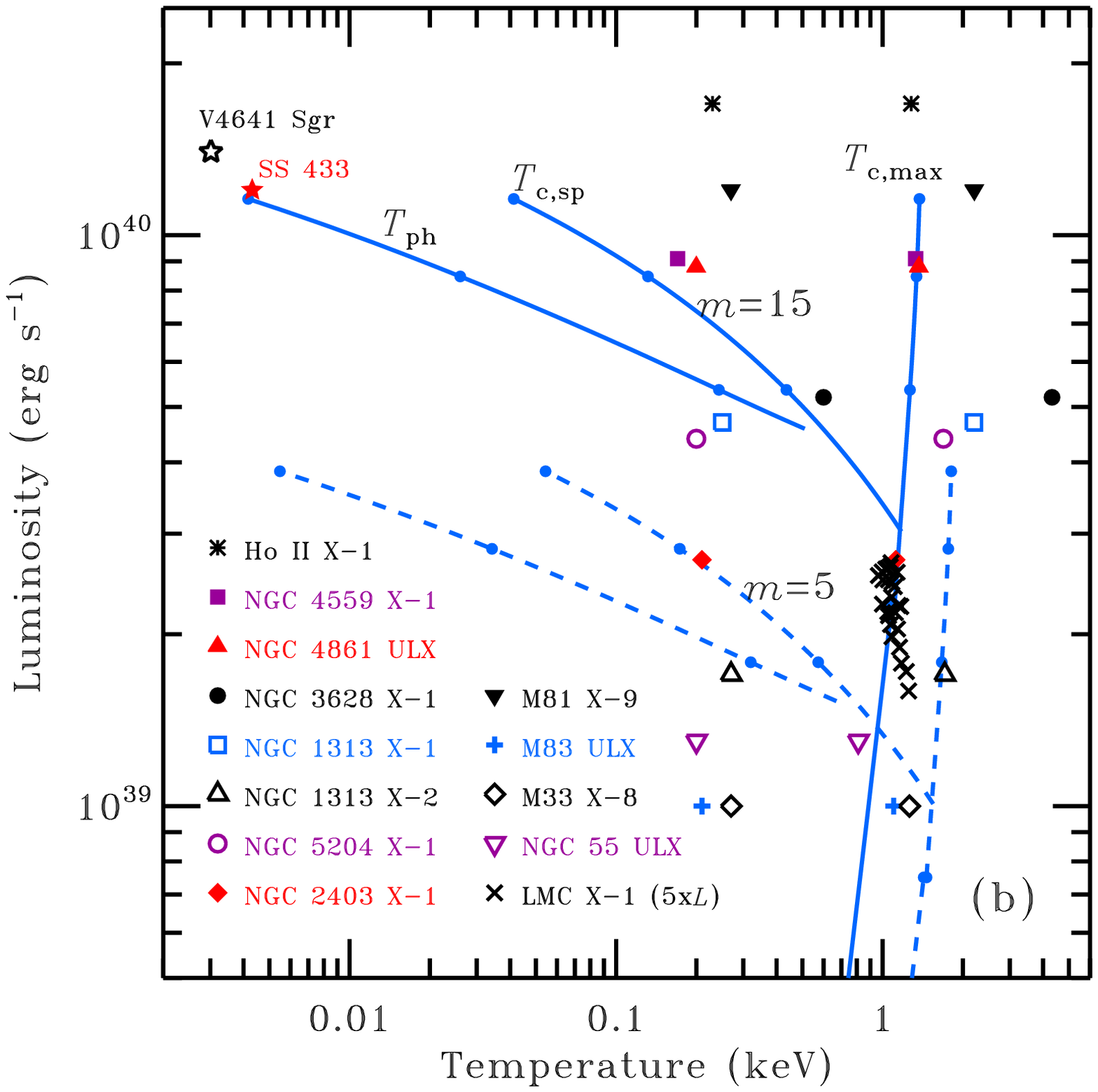}
\end{center}
\caption{\label{fig:hr}
(a) 
The luminosity-temperature relation for sub-critically accreting BHs in the Milky Way and 
Large Magellanic Cloud   (data from \citealt{GD04}). 
Most objects (except LMC X-1) show a well pronounced correlation $L\propto \Tmax^4$ 
consistent with the standard accretion theory (SS73). 
The bolometric luminosity and the color temperature $\Tcmax$ are corrected here 
for the effect of inclination and relativistic effects \citep[see details in][]{GD04}.
Theoretical dependences are shown for 1.5, 5 and 10 solar mass BHs. 
(b) The luminosity-temperature relation for super-critically accreting BHs. 
The curves are theoretical dependences for the model of the advective disc with the
outflow given by  equations  (\ref{eq:ltot}) and (\ref{eq:tmax})--(\ref{eq:Tphout}). 
The accretion rate is $\mdoto\approx2$ at the point of separation of 
various temperatures, and the curves continue until $\mdoto=10^{3}$ 
(filled circles at the curves indicate a change in $\mdoto$ by a factor of 10). 
The following temperatures 
are shown (from the right curve to the left): the maximal color disc temperature $\Tcmax=\fc\Tmax$ with
the color correction factor $\fc=1.7$, 
the color temperature at the spherization radius $\Tcsph$ (also with $\fc=1.7$), 
and the temperature at the outer photosphere $\Tph$.
The wind parameters are $\beta=\zeta=1$, $\epsilonw=1/2$.
The upper (solid) set of curves is for the BH mass $m=15$ 
and the lower dashed curves are  for $m=5$.
The stars show the positions of the super-critically accreting stellar-mass BHs, 
SS 433 \citep{DF97} and V4641 Sgr  \citep{RGC02}.
Other symbols are the  apparent bolometric luminosities and the 
temperatures obtained from the spectral fits
with the blackbody and {\sc diskbb} model to the {\it XMM-Newton} data 
of a set of ULXs (from Table 5 of \citealt{SRW06}). 
The crosses show the data for LMC X-1 if one increases the apparent luminosity by 
a factor of 5.
}
\end{figure*}

 \subsection{Photospheres and the emitted spectrum}
 
We can define three characteristic radii (see Fig.~\ref{fig:tau}): (1) 
the radius of the inner photosphere $\rphin$, where $\tau_{\perp}=1$; (2) the spherization radius 
$\rsph \approx \mdoto$, where the optical depth through the wind in the normal direction is maximal; 
and (3) the outer photosphere $\rphout$, where the wind becomes transparent  
$\tau_{\perp} \approx \tau_{\parallel}=1$.

The inner photosphere is almost independent of $\mdoto$:  
\be \label{eq:rphin}
\rphin \approx  1 + \frac{\beta}{3\epsilonw} . 
\ee
The outer photosphere (for  $\mdoto\gg1$),
\be \label{eq:rphout}
\rphout \approx 3 \ \frac{ \epsilonw}{\zeta\beta} \ \mdoto^{3/2} , 
\ee
is much larger than the spherization radius.

A face-on observer  would see the emission from 
three separate zones defined by the 
three characteristic radii:
\be  \label{eq:zones}
\begin{array}{ll}
r<\rphin, & \mbox{zone A} , \\
\rphin <r<\rsph, & \mbox{zone B} , \\
\rsph <r<\rphout, & \mbox{zone C} . 
\end{array} 
\ee
The characteristic disc temperatures can be obtained from the 
Stefan-Boltzmann law
\be \label{eq:q_ab}
\Qrad(R)= \sigmasb T^4 (R) .
\ee
For the advective disc with $\epsilonw=1/2$ the maximum effective temperature is 
about
\be \label{eq:tmax}
\Tmax = 1.6 \ m^{-1/4} \left( 1 - 0.2 \mdoto^{-1/3} \right) \ \mbox{keV}.
\ee
It  is reached at $r_{\max}\approx 1.06<\rphin$ for large $\mdoto$ and varies little
with the accretion rate. Variations in $\epsilonw$ affects this temperature only 
by a per cent or so (see Fig. \ref{fig:qradmdot}a). 
The observed color temperature differs from $\Tmax$ by a color correction factor $\fc$.
The exact values for $\Tmax$ and $r_{\max}$, however,  
depend on general relativity corrections, which are neglected here.
In zone A, the wind is transparent (i.e. it is momentum-driven) 
and the radiation escapes unaffected by the outflow. 

In  zone B, the wind is opaque and the energy generated in the disc 
is advected by the wind. The ratio of the photon diffusion time in the wind,
$\tau_{\perp} \zeta R/c$, to the dynamical time-scale, $\zeta R/v_{\rm w}$,
is $3\epsilonw/\sqrt{6}\sim 1$ (this supports the view that the wind here is energy-driven).  
Thus  the radiation escapes at a radius about twice the energy generation radius. 
This does not change the radial 
dependence of the effective temperature $T\propto R^{-1/2}$,
resulting in a power-law spectrum $F_{E}\propto E^{-1}$ extending 
from about $\Tphin$ to the temperature at Êthe spherization radius $\rsph\approx\mdoto$:
\be \label{eq:tsph}
\Tsph\approx 1.5 \ m^{-1/4} \mdoto^{-1/2}  \left(1 + 0.3 \ \mdoto^{-3/4}\right)\ \mbox{keV} .
\ee
The resulting temperature should  also be reduced by $(1-\epsilonw)^{1/4}$,
because some energy is transferred to the outflow.

The outer zone C emits about the Eddington luminosity which 
is produced mostly in the disc at radii $r>\rsph$. The photon diffusion time here 
is smaller than the dynamical time, thus most 
of the radiation escapes not far from the radius it is produced. This results in 
the effective temperature variation close to $r^{-3/4}$ 
and the nearly standard spectrum $F_E\propto E^{1/3}$ (SS73).

An edge-on observer would see only the blackbody-like emission 
corresponding to the temperature at the outer photosphere, which 
for $\mdoto\gg1$ (i.e. $\rphout\gg\rsph$) takes the   form:
\be \label{eq:Tphout}
\Tphout \approx 0.8 \left( \frac{\zeta\beta}{\epsilonw} \right)^{1/2}
 m^{-1/4} \mdoto^{-3/4}  \ \mbox{keV} . 
\ee
For accretion rates slightly exceeding the Eddington, $\rphout\lesssim\rsph$, and
 the dependence of $\Tphout$ on  $\mdoto$ is much stronger. 

At intermediate inclinations, 
the central hot part of the disc may be partially blocked by the wind, and 
an observer would see a soft spectrum peaking at $\Tsph$.

\section{Comparison with observations}
\label{sec:obs}

The standard model for sub-critically accreting BHs (SS73;   Sect. \ref{sec:ss73}) 
predicts the relation $L\propto \Tmax^4 \propto \Mdot$. 
At super-Eddington accretion rates,   three characteristic temperatures are 
identified: (i) the maximal color disc temperature 
$\Tcmax=\fc\Tmax \approx 1.6 \fc m^{-1/4}$ keV,  
 (ii) the color temperature at the spherization radius   
 $\Tcsph \approx 1.5 \fc  m^{-1/4} \mdoto^{-1/2}$ keV, 
 and (iii) the outer photosphere temperature  given by equation (\ref{eq:Tphout}). 
 The bolometric luminosity  
and the temperatures depend parametrically on $\mdoto$ 
 according to equations (\ref{eq:ltot}) and (\ref{eq:tmax})--(\ref{eq:Tphout}). 
 These  theoretical dependences are shown in Fig. \ref{fig:hr}(b). 
The luminosity observed  along the symmetry 
axis  may exceed $\Lbol$ by a factor  $1/(1-\cos\theta)$$\sim$2--10 
for the outflow height $z/R$$\sim$0.6--2. 
On the other hand, a fraction  $\epsilonw$ is spent on acceleration of the outflow. 
Together these effects result in the 10--30-fold excess over $\Ledd$ at high accretion rates
\citep[see also][]{OM05,BKP06,FK06}. 
Thus the absolute maximum apparent luminosity in our model 
is about $10^{41}\erg\ \secinv$ for a 20 $\msun$ BH. 

The BHs in the Milky Way and LMC, accreting at a rate above a few per cent 
of Eddington, show spectra peaking at 0.4--1.5 keV. They closely follow 
 the standard  $L\propto \Tmax^4$ dependence   \citep[except LMC X-1, see][ and Fig. \ref{fig:hr}a]{GD04}.
The spread  around it can result from varying contribution of the non-thermal emission.  
A shift of the microquasar GRO J1655-40 to the right from the $L$--$T$ relation for a 7 $\msun$ BH 
\citep{OB97} is probably related to a high spin of the BH there, which results in a higher disc 
temperature compared to the Schwarzschild BH.

Two  super-Eddington accretors in our galaxy, a persistent source 
SS 433 \citep{DF97} and a super-critical transient V4641 Sgr \citep{RGC02}, 
show optical spectra with characteristic temperatures of 30\ 000--50\ 000 K, which we 
associate with the emission from the outer photosphere (and not 
with the spherization temperature $\Tsph$ as proposed by \citealt{BKP06}). 
We estimate the accretion rate to be about $10^3$ higher than the Eddington one.
SS 433 is  a bright UV source with  total luminosity of about $10^{40}\erg\ \secinv$
\citep{DF97}, but is underluminous in the X-rays. 
V4641 Sgr, on the other hand,  exceeded 12 Crab in the X-rays during the outburst and 
and was extremely bright in the optical band. 
Such a difference may be caused by a different mode of accretion 
and/or different inclination of the two systems.

In Fig. \ref{fig:hr}(b) we present the data for ULXs with the luminosities exceeding 
$10^{39}\erg\ \secinv$  and the most reliable spectra obtained by {\it XMM-Newton}.
The temperatures obtained by \citet{SRW06} 
from the fits with the blackbody and {\sc diskbb} models and the 
corresponding observed bolometric luminosities are plotted.   
A theoretical spectrum from a super-critical disc with 
a $T(R)\propto R^{-1/2}$ dependence, can, in principle,   be represented as a sum of such components. 
We associate a high temperature component with the 
hot inner disc of  $\Tmax$$\sim$1 keV. 
A high BH spin and an overheating of the disc above the effective temperature 
at $\mdoto$$\sim$2--20 \citep{B98,Sul02,Kaw03} may be responsible for sometimes
observed higher (up to 4 keV) temperatures.
A soft, $\sim$0.2 keV component may correspond 
to  the spherization temperature implying the 
accretion rate $\mdoto= m^{-1/2}(1.5\fc/\Tcsph [\mbox{keV}])^2$$ \approx$30--40 onto a 
stellar mass, 10--20 $\msun$, BH. The observed higher luminosities 
can result from the geometrical beaming. 

LMC X-1 deserves a more detailed discussion.  
Its apparent luminosity is sub-Eddington for a 10 $\msun$ BH, but its 
track on the $L$--$T$ diagram (see Fig. \ref{fig:hr}a) is perpendicular to that for other BHs.
This can  be understood if its mass  is only $\sim$1.5$\msun$ 
\citep[i.e. it can be a neutron star, but see][]{HCC83,E91} 
and the source accretes at $\mdoto$$\sim$3. 
The observed rather soft spectrum is associated with the spherization region and results from 
obscuration of the central hot disc. In that case, however, we  seriously 
underestimate the true luminosity of the object. 
Assuming that it is five time larger than the apparent one, 
the behaviour of LMC X-1 on the $L$--$T$ plane  
(Fig. \ref{fig:hr}b) becomes consistent with that 
of the spherization temperature at $\mdoto\approx$1--3 for a 10 $\msun$ BH.

\section{Summary} 

A black hole accreting at a super-Eddington rate is likely to produce 
strong winds (and jets) as observed in SS 433, the only known BH in the Milky Way 
accreting persistently at such a high rate. The bolometric luminosity  
can exceed the Eddington limit by a logarithmic factor $\sim \ln\mdoto$, 
while the apparent luminosity can be factor of 5 higher because of the geometrical beaming. 
The observational appearance of such an object strongly depends on the inclination 
angle of the system to the line-of-sight. In edge-on systems, the central X-ray source 
is hidden by the wind and most of the radiation, with  the characteristic temperature of 
$\sim 10^4$--$10^5$ K, escapes in the UV band. 
A face-on observer sees the hot inner flow with the flat in $EF_E$ spectrum extending 
from a few keV down to the temperature at spherization radius $\Tsph\approx \mdoto^{-1/2}$ keV. 

Strong winds from the accretion disc 
explain naturally the presence  around the ULXs of the  expanding nebulae
which are photoionized by the X-ray and UV radiation of the central source.
An excellent agreement between the model and the data
supports views \citep{FM01,K02,Fab04,BKP06,VM06} 
that ULXs are super-critically accreting stellar-mass BHs
similar to SS 433, but observed along the symmetry axis.

\section*{Acknowledgements}
 This work was supported by the Academy of Finland grants  
 102181, 111720, 107943,  109122, 110792 and 112982,
the Magnus Ehrnrooths Foundation, the
Vilho, Yrj\"o and Kalle V\"ais\"al\"a Foundation, 
the Russian RFBR grants 04-02-16349  and 06-02-16025-a, 
and the RFBR/JSPC grant 05-02-19710. 
JP thanks the Kavli Institute for Particle Astrophysics and 
Cosmology and the Max-Planck-Institut f\"ur Astrophysik 
for hospitality during his visits.
We thank Marek Gierli\'nski for the data and 
Ramesh Narayan for valuable suggestions. 



\label{lastpage}

\end{document}